# All-Inorganic p−n Heterojunction Solar Cells by Solution Combustion Synthesis using n-type FeMnO$_3$ Perovskite Photoactive Layer

**Ioannis T. Papadas,[a,c]\* Apostolos Ioakeimidis,[a] Ioannis Vamvasakis,[b] Polyvios Eleftheriou [a] Gerasimos S. Armatas[b] and Stelios A. Choulis[a]\***

[a] Molecular Electronics and Photonics Research Unit, Department of Mechanical Engineering and Materials Science and Engineering, Cyprus University of Technology, Limassol, Cyprus.
[b] Department of Materials Science and Technology, University of Crete, Heraklion 70013, Greece.
[c] Department of Public and Community Health, School of Public Health, University of West Attica, Athens, Greece

**\* Correspondence:**
Corresponding Authors: Assistant Prof. Ioannis T. Papadas, Prof. Stelios A. Choulis
emails: ioannis.papadas@cut.ac.cy, stelios.choulis@cut.ac.cy

## Abstract

This study outlines the synthesis and physicochemical characteristics of a solution-processable iron manganite (FeMnO$_3$) nanoparticles via a chemical combustion method using tartartic acid as a fuel and demonstrates the performance of this material as a n-type photoactive layer in all-oxide solar cells. It is shown that the solution combustion synthesis (SCS) method enables the formation of pure crystal phase FeMnO$_3$ with controllable particle size. XRD pattern and morphology images from TEM confirm the purity of FeMnO$_3$ phase and the relative small crystallite size (~13 nm), firstly reported in the literature. Moreover, to assemble a network of connected FeMnO$_3$ nanoparticles, β-alanine was used as a capping agent and dimethylformamide (DMF) as a polar aprotic solvent for the colloidal dispersion of FeMnO$_3$ NPs. This procedure yields a ~500 nm thick photoactive layer. The proposed method is crucial to obtain functional solution processed NiO/FeMnO$_3$ heterojunction inorganic photovoltaics. The optoelectronic properties of the heterojunction were established. These solar cells demonstrate a high open circuit voltage of 1.31 V with sufficient fill factor of 54.3% and low short circuit current of 0.07 mA cm$^{-2}$ delivering a power conversion efficiency of 0.05% under 100 mW cm$^{-2}$ illumination. This work expands on the burgeoning of environmentally friendly, low-cost, sustainable solar cell material that derive from metal oxides.

## 1. Introduction

A major source of renewable energy is solar energy.[1] Nevertheless, the production of fuels and electricity from solar power is still costly, mainly because of the materials used in building the cells.[1,2] Except for Si and copper indium gallium selenide solar cells (CIGSSe), which are principal targets for photovoltaic (PV) applications, CdTe and GaAs are also significant photoactive materials. However, the mass production of such photovoltaics is limited due to the high production costs, the indirect bandgap energy (for Si) and the dependence on elements which are expensive (In, Ga, Te) or even hazardous (As, Cd).[3–6] Contrarily, hybrid lead halide perovskite solar cells (PVSCs) have been the epicenter of the current solar cell research because of their facile fabrication process and due to the fact that they appear a high PCE of over 23%.[7] However, PVSCs have disadvantages as well, that is, ultraviolet light absorption, and humidity and atmospheric oxygen affecting decomposition.[8,9] Along with the presence of toxic lead, these factors restrain the advancement of the PVSCs. Therefore, it is essential to find alternatives for inorganic PVs that use inexpensive, eco-friendly and earth abundant photoactive materials with appropriate semiconductor structure, while at the same time seek improvements in solar cell efficiency.

Metal oxide (MeOx) based solar cells have the potential to resolve some of the issues which arise in conventional solar cells. The all-oxide perspective is advantageous due to its excellent chemical stability, minor toxicity and ample quantities of metal oxides that effectively permit the manufacturing of solar cells under ambient conditions.[10] MeOx are typically used as functional layers in solar cells such as transparent conducting front electrodes (ITO, FTO),[11,12] electron ($TiO_2$, $SnO_2$, ZnO, $Fe_2O_3$ etc.)[13–16] or hole (Cu:NiOx, $CuGaO_2$, $NiCo_2O_4$, CuLi:$NiCo_2O_4$ etc.) transporting layers,[17–20] whereas a very small number of MeOx have been used as photoactive layers, primarily $Cu_2O$, CuO and $Co_2O_3$.[10]

Ferroic semiconductors are now being continually included in the list of materials which are employed to investigate and push the efficiency limits in all-oxide photovoltaics. Ferroic semiconductor materials are considered to be light absorbers, such as Pb(Zr,Ti)$O_3$ (energy band gap ($E_g$) = 3.63 eV),[21] $KNbO_3$ ($E_g$ = 3.8 eV)[22] and $BiFeO_3$ ($E_g$ = 2.67 eV)[23]. However, what seems to confine their implication to solar cells is their wide energy bandgap that results to low absorption of visible light and, thus, low conductivity. Other ferroelectric oxide semiconductors, like $Bi_2FeCrO_6$[24] and $BiMnO_3$[25], have a suitable bandgap between 1–2 eV and are considered as more efficient in absorbing solar light. It is worth mentioning that a power conversion efficiency (PCE) up to ~8.1 % has been recently achieved using a single ferroelectric $Bi_2FeCrO_6$ layer fabricated by pulse laser deposited technique with the following structure: $SrTiO_3$/$SrRuO_3$/$Bi_2CrFeO_6$/ITO.[24] However, without the mentioned method of the thin film deposition which is considered to be a very complicated and energy demanding procedure, it is very unlikely to gain high PCE values.[26] Furthermore, the usage of the low-bandgap $KBiFe_2O_5$ material ($E_g$ ~1.6 eV) in the photovoltaic cells appears to be restricted due to low PCE (~3 $10^{-3}$ %).[27] Therefore, the above mentioned examples demonstrate that further research is needed in finding more efficient narrow-bandgap, non-toxic and low-cost materials for solar cell devices.

The sol-gel synthesis has been the most frequently used method in the manufacture of MeOx. Nevertheless, to achieve crystallinity and to guarantee an efficient charge–carrier mobility, for metal oxide based active layers, high temperatures are necessary, which increases the cost of manufacturing and also restricts printable applications. These restrictions call for alternative techniques which could operate at lower temperatures. Compared to sol-gel synthesis of MeOx nanoparticles (NPs), the solution combustion synthesis (SCS) of NPs displays considerable advantages, such as use of a simple experimental setup, production of NPs with high crystallinity and pure phase, and exact control of the size and crystal structure of the particles by simple adjusting the fabrication conditions.[28–34] The SCS seems to be adaptable and effective for the growth of high crystalline MeOx layers at relative lower temperatures. As an exothermic procedure with a high rate of heat release, the necessity for high temperatures is circumvented and high purity MeOx NPs can be produced at moderate reaction conditions. In SCS process, the metal salts (e.g., nitrates) dissolved in saturated aqueous or alcoholic solutions act as oxidizing agents and react with organic fuels (such as urea, glycine, acetylacetonate, citric acid etc.) under relatively lower temperatures compared to other commonly used solution process methods to give rise to a combustion reaction and to produce the corresponding metal oxide NPs.[35–37]

Iron manganite (FeMnO$_3$, FMO) is a mixed perovskite material with the chemical formula ABO$_3$, where the Fe atom is placed at the center of a cube formed by eight corner-sharing MnO$_6$ octahedra.[38] FeMnO$_3$ has been examined for applications such as lithium-ion batteries, catalysis, humidity sensors, energy storage and antibacterial devices.[39–43] A large number of synthesis methods, such as co-precipitation, hydrothermal, ball milling, solid state reaction and sol-gel chemistry, have all been employed for the fabrication of FeMnO$_3$ materials.[39-51] Despite this, not all these techniques are viable to synthesize FeMnO$_3$ nanomaterials, as there are some drawbacks such as the expense of the source materials, chemical non-uniformity, high impurity, aggregated nanoparticles, and non-stoichiometry of some ferrite systems.[52–55] FeMnO$_3$ is a semiconductor which consists of plentiful and environmentally friendly elements with an ideal direct optical bandgap (~1.5 eV) to absorb solar photons, while it has a deep lying valence band (VB ~ 5.3 eV)[52] that corresponds well with the VB edges of several p-type materials (e.g., CuO, NiCo$_2$O$_4$ etc.).[19,56] Additionally, it has high photochemical stability which is necessary for long-term optoelectronic devices and its intrinsic electric polarization field can enable charge-carrier separation within the semiconducting structure. Such characteristics make FeMnO$_3$ a promising light absorber for optoelectronic uses. On the other hand, amongst the various metal oxides that have been used as p-type active layers, nickel oxide (NiO) is a promising candidate for PVs due to its excellent electrochemical behavior.[57] NiO has a rock salt structure and exhibits high p-type conductivity with a wide bandgap between 3.5 and 4.0 eV, making it transparent to the visible light.[58–60]

In this work, SCS of FeMnO$_3$ NPs is presented and indicating that tartaric acid can be used as a fuel and nitrate as an oxidizer agent. FeMnO$_3$ NPs with an average size of ~13 nm and narrow particle-size distribution were prepared using a low cost SCS process (6 h calcination at 450 °C in air). The as-synthesized FeMnO$_3$ NPs were then functionalized with $\beta$-alanine and the ligand-capped NPs enabled the formation of compact and functional layers. These films were used, for the first time, as n-type photoactive materials and were incorporated in p-n heterojunction of all-oxide solar cells. For the purposes of this study, nanostructured NiO films were also synthesized by SCS method and applied as a p-type layer in the following structure ITO/NiO/FeMnO$_3$/Cu. The corresponding MeOx PVs show a high $V_{oc}$ of 1.31 V with adequate FF of 54.3% and limited short current of 0.07 mA cm$^{-2}$ resulting to a PCE of 0.05%. Electrical characterizations by impedance spectroscopy reveled a high charge recombination resistance inducing high $V_{oc}$, whereas the limited current density is ascribed to the high charge transport resistance. Despite the low PCE values, these results provide a framework for further optoelectronic properties research on eco-friendly and cost-effective photoactive layers for fabrication of all solution processable inorganic photovoltaics.

## 2. Experimental
*Materials:* Pre-patterned glass-ITO substrates (sheet resistance 4Ω/sq) were purchased from Psiotec Ltd. All the other chemicals used in this study were purchased from Sigma Aldrich.
*Solution combustion synthesis (SCS) of FeMnO$_3$ NPs:* For the synthesis of FeMnO$_3$ NPs, 0.5 mmol Mn(NO$_3$)$_2$.4H$_2$O, 0.5 mmol Fe(NO$_3$)$_3$.9H$_2$O and tartaric acid were blended in 5 ml of 2-methoxy ethanol solution. Subsequently, 150 μL HNO$_3$ (69% wt HNO$_3$) were added slowly into the mixture, and the solution stirred up to almost complete homogeneity. The whole solution was left under stirring for at least 3 h at room temperature (RT). The molar ratio of the total metal nitrates and tartaric acid was 1. Thereafter, the precursor solution was heated at 100 °C under consecutive stirring until complete evaporation of the solvent. The dry black powder was then used for the combustion synthesis of the FeMnO$_3$ NPs in ambient atmosphere at 450 °C in a preheated oven for 6 h, so that the combustion process be completed and then left to cool down at room temperature.

*Perovskite FeMnO$_3$ films preparation:* The prepared FeMnO$_3$ (FMO) NP powder was used for the preparation of FMO dispersion for the deposition of corresponding film by spin coating technique. Firstly, the surface of NPs was modified with *β*-alanine. Briefly, as-made FMO NPs (50 mg) were added in 4 mL of deionized (DI) water containing *β*-alanine (10 mg), and the pH of the solution was adjusted to 4.2 with 1M HNO$_3$.[52,61] To secure that NPs will transfer to the liquid phase and form a stable suspension, typically within 1 day, the resulting mixture was then intensively stirred at RT. The dispersion was aided with probe sonicator for about 30 min. To a stable colloidal dispersion of 30 mg mL$^{-1}$ be formed, the alanine-capped FMO NPs were isolated by centrifugation, rinsed several times with DI water, and finally dispersed in DMF. The obtained homogenous dispersion was then drop-casted and subsequently was spin coated on the top of NiO layer at 3000 rpm for 40 sec. The process of FeMnO$_3$ film formation was repeated about ten times to obtain a desired thickness of about 500 nm. To assemble a network of tightly connected metal oxide NPs, the deposition of the FeMnO$_3$ films was accomplished by spin coating technique of the colloidal NPs, followed by thermal annealing at 300 °C for 30 min. In this way, *β*-alanine can enable direct NP–NP interactions upon ligand removal at growth temperature due to its small size, thus yielding high strength films consisted of firmly interconnected NP networks. This strategy is incredibly important to obtain functional metal oxides photovoltaic devices with good charge transfer properties.

*NiO NPs synthesis and films preparation by SCS:* For the solution combustion synthesis of NiO, 1 mmol of Ni(NO$_3$)$_2$·6H$_2$O were dissolved in 2.5 ml of 2-methoxyethanol. After the solution was stirred at 50 °C for 1 h, 0.1 mmol of acetylacetone was added to the solution, and the whole solution was allowed under further stirring for 1h at RT. Spin coating technique was applied for the fabrication of the precursor films on the various substrates. The precursors' solution was spin coated at 3000 rpm for 40 s. The resulting light green colored films were dried at 100 °C for 5 min and used as a precursor for the combustion synthesis of NiO NPs. Subsequently the obtained films were heated at 300 °C in ambient atmosphere for 1 h in a preheated hot plate to complete the combustion process and then left to cool down at room temperature, forming a ~50 nm thin layer.

*Device fabrication:* The metal oxides solar cells under study were ITO/NiO-NPs/FeMnO$_3$-NPs/Cu. ITO substrates were sonicated in acetone and subsequently in isopropanol for 10 min and then heated at 100 °C on a hot plate for 10 min before use. The substrates were further treated with ozone for 10 min to achieve a better contact with the active layer by reducing the contact resistance. To fabricate the devices, a layer of NiO as p-type and FeMnO$_3$ as n-type side of the p-n junction were formed in sequence. The deposition of corresponding metal oxides films was described in detail above. Finally, 200 nm Cu layers were thermally evaporated through a shadow mask to finalize the devices, giving an active area of 0.9 mm$^2$.

*Characterization:* Thermogravimetric Analysis (TGA) were performed on a Shimadzu Simultaneous DTA-TG system (DTG-60H). Thermal analysis was conducted from 40 to 600 °C in air atmosphere using air gas with a flow rate of 200 mL min$^{-1}$ and a heating rate of 10 °C min$^{-1}$.

X-ray diffraction (XRD) patterns were collected on a PANanalytical X´pert Pro MPD powder diffractometer (40 kV, 45 mA) using Cu Kα radiation (λ=1.5418 Å). Transmission electron microscope (TEM) images and electron diffraction patterns were recorded on a JEOL JEM-2100 microscope with an acceleration voltage of 200 kV. The samples were first gently ground, suspended in ethanol, and then picked up on a carbon-coated Cu grid. Quantitative microprobe analyses were performed on a JEOL JSM-6390LV scanning electron microscope (SEM) equipped with an Oxford INCA PentaFET-x3 energy dispersive X-ray spectroscopy (EDS) detector. Data acquisition was performed with an accelerating voltage of 20 kV and 60 s accumulation time. Absorption measurements were performed with a Schimadzu UV-2700 UV-Vis spectrophotometer. For UV-VIS and PL measurements, thick films of $FeMnO_3$ NPs have been fabricated on top of the quartz substrates employing the spin coating method. UV–vis/near-IR diffuse reflectance spectra were recorded with a Schimadzu UV-2700 UV-Vis spectrophotometer, using $BaSO_4$ powder as a 100 % reflectance standard. The energy bandgap ($E_g$) of the samples were estimated from Tauc plots of $(Fh\nu)^2$ as a function of photon energy ($h\nu$), where F is the Kubelka–Munk function of the reflectance (R): $F=(1-R)^2/(2R)$.[62] The thickness of the films were measured with a Veeco Dektak 150 profilometer. The PL measurements were performed on $FeMnO_3$ film on quartz substrate at an excitation wavelength of 400 nm. Photoluminescence (PL) spectrum was obtained at room temperature on a Jobin-Yvon Horiba FluoroMax-P (SPEX) spectrofluorimeter (Singapore) equipped with a 150 W Xenon lamp and operated from 300 to 900 nm. The current density-voltage (J-V) characteristics were characterized with a Botest LIV Functionality Test System. Forward bias scans were measured with 10 mV voltage steps and 40 msec of delay time. For illumination, a calibrated Newport Solar simulator equipped with a Xe lamp was used, providing an AM1.5G spectrum at 100 mW/cm$^2$ as measured by a certified oriel 91150V calibration cell. A shadow mask was attached to each device prior to measurements to accurately define the corresponding device area. EQE measurements were performed by Newport System, Model 70356_70316NS. Atomic force microscopy (AFM) images were obtained using a Nanosurf easy scan 2 controller under the tapping mode. Electro-impedance spectroscopy (EIS) and Mott-Schottky measurements were performed using a Metrohm Autolab PGSTAT 302N, where for the EIS a red light-emitting diode (LED) (at 625 nm) was used as the light source calibrated to 100 mW/cm$^2$. For EIS a small AC perturbation of 20 mV was applied to the devices, and the different current output was measured throughout a frequency range of 1 MHz-1 Hz. The steady state DC bias was kept at 0 V throughout the EIS experiments. Mott-Schottky measurements on $FeMnO_3$ films were performed in a 0.5 M $Na_2SO_4$ aqueous electrolyte (pH = 7) using a Metrohm Autolab PGSTAT 302N potentiostat. A three-electrode set-up, with a platinum plate (1.0 × 2.0 cm$^2$) and a silver-silver chloride (Ag/AgCl, 3M KCl) as the counter and reference electrodes, respectively, was adopted to study the samples. The capacitance of the semiconductor/electrolyte interface was obtained at 1 kHz, with 10 mV AC voltage perturbation. All the experiments were conducted under dark conditions. The measured potential vs the Ag/AgCl reference electrode was converted to the normal hydrogen electrode (NHE) scale using the formula: $E_{NHE} = E_{Ag/AgCl} + 0.210$ V. The working electrode for impedance-potential measurement was fabricated as follows, 10 mg of $FeMnO_3$ NPs was dispersed in 1 mL DI water and the mixture was subjected to sonication in a water bath until a uniform suspension was formed. After that, 100 µL of the suspension was drop-casted onto the surface of fluorine-doped tin oxide (FTO, 9 Ω/sq) substrate, which was masked with an epoxy resin to expose an effective area of 1.0 × 1.0 cm$^2$. The sample was dried in a 60 °C oven for 30 min.

## 3. Results and Discussion

The solution processing technique provides an extensible low cost deposition procedure to fabricate high quality metal-oxide films and to replace costly and time consuming vacuum-deposition methods.[63] SCS has recently been employed in the low temperature manufacture of spinel nickel cobaltite ($NiCo_2O_4$) thin films as hole transport layers (HTLs) in inverted p-i-n perovskite solar cells.[19] SCS possesses the benefit of speedily creating homogenous metal oxide materials of fine grain size and, most notably, at a lower temperature than the standard solid–state reaction, sol-gel and co-precipitation techniques.[64,65] The metal nitrates are distinguished for their capacity to synthesize metal oxides films of superior quality. Furthermore, it is important to choose the appropriate fuel agent for combustion, so as to circumvent the creation of sizeable clusters and particle agglomeration.[30,66] Tartaric acid was therefore used as the fuel agent in this work as it results in the formation of a single-crystalline phase of $FeMnO_3$. In general, tartaric acid leads to the formation of stable heterometallic polynuclear complexes[67] because of its carboxylate and hydroxyl groups which can bind different metal ions from the solution, such as $Mn^{2+}$ and $Fe^{3+}$.[68] Basically, the growth of $FeMnO_3$ NPs is the result of combustion reaction of these polynuclear complexes while being heated in the presence of concentrated $HNO_3$.[63,69]

### 3.1. Synthesis and characterization $FeMnO_3$ nanoparticles

To achieve the solution combustion synthesis of $FeMnO_3$ NPs, tartaric acid and metal nitrates precursors were dissolved in 2-methoxyethanol. The precursor's solution was heated at 100 °C under stirring until complete evaporation of the solvent. The obtained gel product was then analyzed by thermogravimetric analysis (TGA). The thermal behavior of the Mn/Fe-tartaric compound was observed by TGA, employing a heating rate of 10 °C min$^{-1}$ in ambient air. As illustrated in Fig. 1a, the reaction shows an acute sudden mass loss at ~230 °C, noted in TGA curve, which is associated with a strong exothermic release of energy during the combustion process. In our study, the as-synthesized material was crystallized well to the perovskite phase in ambient atmosphere at 450 °C in a preheated oven for 6 h.

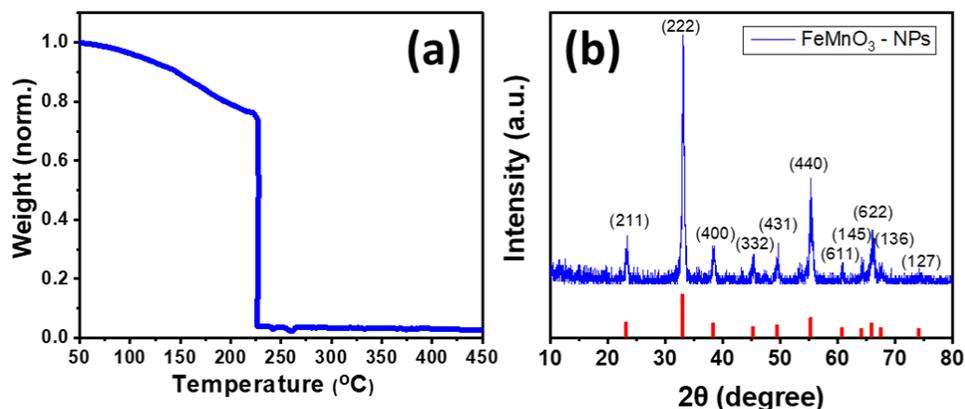

**Fig. 1** (a) TGA profile of the as-prepared $FeMnO_3$ material synthesized via solution combustion process. The weight loss was normalized to the initial sample mass. (b) XRD pattern of $FeMnO_3$ NPs obtained at 450 °C annealing temperature.

The XRD measurement confirmed the crystallinity and phase purity of the $FeMnO_3$ NPs produced through the SCS method. Fig. 1b depicts the XRD pattern of nanocrystalline $FeMnO_3$ obtained at 450 °C soaking temperature. All the diffraction peaks compare well with the reported cubic iron manganite structure (JCPDS card #75-0894) with a=b=c=9.4 Å and α= β=γ=90° unit cell parameters. No peaks from impurity phases, like MnO or $Fe_2O_3$, were observed in XRD pattern, showing the phase purity of the sample. The mean $FeMnO_3$ crystallite size is estimated at ~15 nm using the Scherrer`s equation and peak broadening of the (222) reflection.

TEM corroborated the phase purity of the obtained $FeMnO_3$ NPs. Fig. 2a illustrates a characteristic TEM image of the $FeMnO_3$ sample fabricated at 450 °C. It is seen that this material is made up of tightly connected NPs with an average diameter of 13 ± 2 nm (inset Fig. 2a), which match well to the crystallite size calculated from XRD. The crystal structure of the $FeMnO_3$ was then examined by selected-area electron diffraction (SAED). The SAED pattern recorded from a small area of the $FeMnO_3$ sample (Fig. 2b) indicates a series of broad concentric diffraction rings, which can be assigned to the cubic phase of $FeMnO_3$.[70] In line with XRD results, no other crystal phases were detected by means of electron diffraction. Furthermore, characterization of the composition of $FeMnO_3$ with EDS analysis revealed a Fe:Mn atomic ratio close to 1:1, in agreement with the stoichiometry of $FeMnO_3$ compound (Fig. S1).

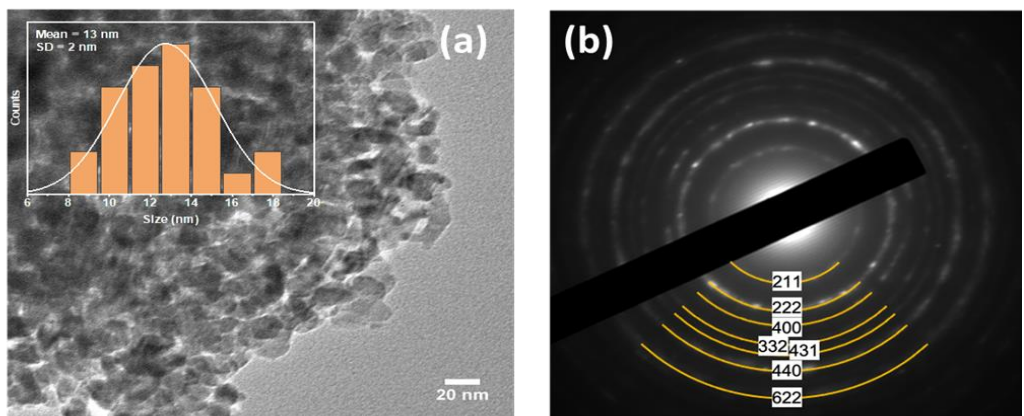

**Fig. 2** (a) Representative TEM image (inset size distribution histogram of the $FeMnO_3$ NPs, showing an average diameter of 13 ± 2 nm), and (b) SAED pattern of the as-synthesized $FeMnO_3$ NPs obtained at 450 °C.

The electronic structure of as-prepared $FeMnO_3$ was also examined by diffuse reflectance ultraviolet-visible/near-IR (UV-vis/NIR) spectroscopy. Fig. 3a shows the UV-vis/NIR absorption spectrum for $FeMnO_3$ NPs synthesized at 450 °C by SCS. This sample shows an acute optical absorption onset in the near IR region (~805 nm), which is associated with an energy gap at ~1.54 eV, as determined by Tauc`s plots [$(Fh\nu)^{1/2}$ versus photon energy (hν), where F, h, and ν are the Kubelka-Munk function of the reflectance, Plank constant and light frequency, respectively],[62] see inset of Fig. 3a.

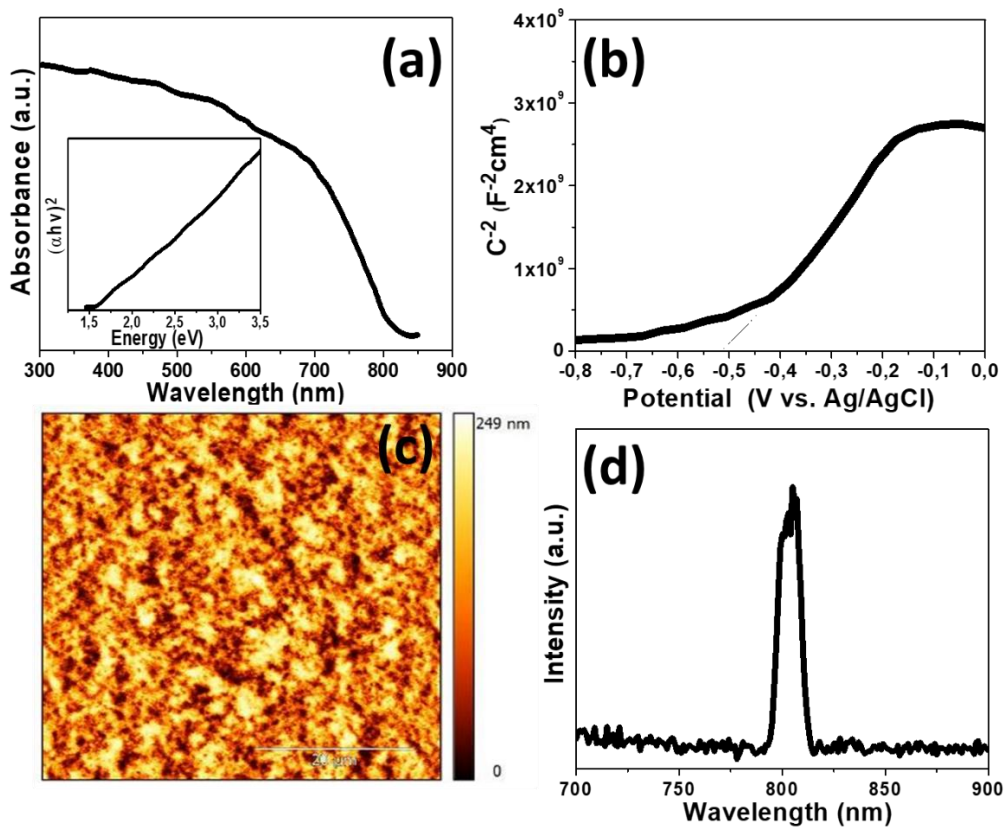

**Fig. 3** (a) UV-Vis/NIR absorption spectrum of FeMnO$_3$ NPs [Inset: the (Fhv)$^2$ versus hv plot derived from optical absorption spectra]. (b) Mott-Schottky plot of the inverse square space-charge capacitance (1/C$_{SC}^2$) as a function of applied voltage (E) relative to the redox potential of Ag/AgCl (3 M KCl) for the FeMnO$_3$ NPs. (c) Typical AFM image of the ITO/FeMnO$_3$ film after SCS synthesis at 450 °C (The scale bar is 20 μm). (d) Room-temperature PL emission spectra of the FeMnO$_3$ NPs.

Electrochemical impedance spectroscopy (EIS) thereafter was employed to examine the position of the conduction band (CB) and valence band (VB) edges of FeMnO$_3$ material. Fig. 3b shows the ensuing Mott-Schottky plot and the matching fit of the linear part of the inverse square space-charge capacitance (1/C$_{sc}^2$) as a function of potential (E). The FeMnO$_3$ reveals a positive linear slope, showing n-type conductivity, where electrons are majority carriers. By using extrapolation to 1/C$_{SC}^2$ = 0, the flat-band potential (E$_{FB}$) of FeMnO$_3$ NPs was estimated to be – 0.31 eV vs NHE (pH=7). Based on the E$_{FB}$ and optical bandgap (as obtained from UV-vis/NIR reflectance data, Fig. 3a) values, the energy band edges for FeMnO$_3$ NPs are CB: –3.78 eV and VB: –5.32 eV vs vacuum.[52] This is further highlighted in the energy level diagram shown in Fig. 4b, which is based on EIS measured values for FeMnO$_3$ and literature data for ITO, NiO and Cu components.[71] For heavily n-typed doped semiconductors, it can be supposed that the E$_{FB}$ level is very close to the CB edge. Generally, for several n-type semiconductors the CB edge is approximately 0.1–0.3 eV higher than the E$_{FB}$ potential. Therefore, the position of the VB edge was estimated from E$_{FB}$ – E$_g$. In iron manganite materials there are a number of reports which connect the electron hopping between $Fe^{+3}$-$Fe^{+2}$ and hole hopping between $Mn^{+2}$-$Mn^{+3}$ ions with n-type and p-type of conductivities.[72,73] The findings of these studies suggest that both n-type and p-type charge carriers are anticipated to contribute to the conduction mechanism in $FeMnO_3$ structure.[74,75] In our study, the positive slope of the Mott-Schottky plots (Fig. 3b) clearly shows that the perovskite iron manganite have n-type behavior.[52]

The smoothness of deposited $FeMnO_3$ film plays a crucial role for the formation of a deep depletion region, which is highly desirable for high performing devices. To achieve this *β*-alanine was used as surface capping ligand for $FeMnO_3$ NPs. In this process, the carboxyl (–$CO_2H$) groups of *β*-alanine can coordinate to the nanoparticles' surface, while the amine (–$NH_2$) functional groups can inhibit the nanoparticles aggregation and stabilize the colloidal solution. This resulted in the formation of a ~500 nm thick compact film (Fig. S2) consisting of a continuous network of tightly interconnected NPs (Fig. S3) with a relative low roughness of ~27 nm, as calculated by AFM topography measurements (Fig. 3c).

For PL measurements of perovskite material, thick films of $FeMnO_3$ NPs were fabricated on top of the quartz substrates employing the spin coating method (for details, see the experimental section). PL spectroscopy is an essential tool for finding the purity and crystalline quality of semiconductors. The PL spectrum of $FeMnO_3$ NPs, in Fig. 3d, shows an intense near band edge emission at ~805 nm. This emission peak corresponds to the CB-VB inter-band transition and no additional peaks due to the radiative relaxations from defect sites or impurities were observed in PL spectrum of $FeMnO_3$.[42]

### 3.2 Photovoltaic device characterization

As a proof of concept, the newly developed $FeMnO_3$ NP aggregates were used as a n-type photoactive material in a p-n full metal oxides solar cell, with the structure ITO/NiO/FMO/Cu (Fig. 4a). Both NiO and $FeMnO_3$ materials were synthesized by the solution combustion method, as it is described in the experimental sections, rendering the fabrication process of such solar cells remarkably facile. Finally, a 200-nm-thick Cu layer was thermally deposited on the surface of $FeMnO_3$ to complete the device (see Fig. 4a).

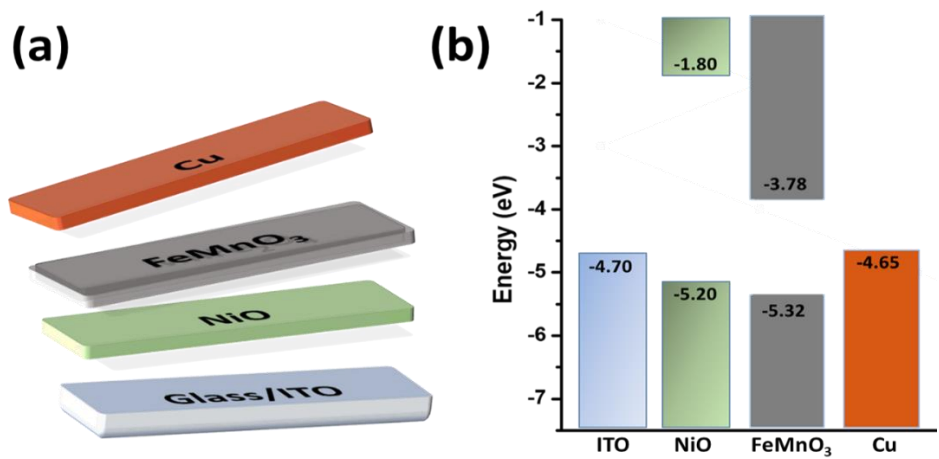

**Fig. 4** (a) Schematic representation, and (b) the corresponding energy level diagram of each component of the ITO/NiO/FeMnO$_3$/Cu device.

Fig. 5a shows the J–V curve of the ITO/NiO/FeMnO$_3$/Cu device under 1 sun simulated light (100 mW cm$^{-2}$) where the curve shape evidence the Schottky barrier formation at the junctions. The extracted photovoltaic parameters, open-circuit voltage (V$_{oc}$), short-circuit current (I$_{sc}$), fill factor (FF), and power conversion efficiency (PCE) are listed in Table 1. The device yields a high V$_{oc}$ of 1.31 V with adequate FF of 54.3%, but the generated current density is low (J$_{sc}$ = 0.07 mA cm$^{-2}$) delivering a PCE of 0.05%.

Among the studied all-oxide ferroic solar cells, that were fabricated by the same solution processes, the highest PCE was obtained by FeMnO$_3$-based solar cell (~0.05 %), which is notably higher than those of the NiO/BiFeO$_3$ (~0.025 %)[76] and pure Pb(ZrTi)O$_3$ (~0.00008 %) based solar cells.[77] Furthermore, the obtained V$_{oc}$ of the FeMnO$_3$-based solar cell (~1.31 V) is also much higher with respect to the solar cells using BiFeO$_3$ (~0.41 V)[76] and Pb(ZrTi)O$_3$ (~0.6 V) as a light absorber, respectively.[77] Therefore, we propose that FeMnO$_3$ could be a promising candidate for solar cell applications.

**Table 1** Extracted solar cell parameters from the J–V characterization of the ITO/NiO/FeMnO$_3$/Cu device

| Solar cell | V$_{oc}$ (V) | J$_{sc}$ (mA/cm$^2$) | FF (%) | PCE (%) |
|---|---|---|---|---|
| ITO/NiO/FMO/Cu | 1.31 | 0.07 | 54.3 | 0.05 |

To examine the spectral response of the device, external quantum efficiency (EQE) measurements were conducted, and the results are presented in Fig. 5b along with the respective integrated photocurrent response. The spectral response of EQE reveals that the photo-generated current is produced in both NiO and FeMnO$_3$ layers with the spectrum correspondingly match the optical absorption spectra of the respective NiO and FeMnO$_3$ films. Specifically, we observe photocurrent generation onsets in the ultraviolent (300–350 nm) and near IR (800–850 nm) regions, which correspond to the acute optical absorption onset of NiO (~355 nm) and FeMnO$_3$ (~805 nm), respectively.[78] Thus, we note that although NiO is a commonly used hole transporting layer, in the device structure under investigation provides a small contribution to the external quantum efficiency as discussed above and for this reason the term NiO/FeMnO$_3$-based heterojunction solar cells is used within the paper. The integrated photocurrent density (0.064

mA/cm$^2$) is also in very close agreement with the value obtained from the J–V curve (0.07 mA/cm$^2$) obtained from the solar simulator analysis.

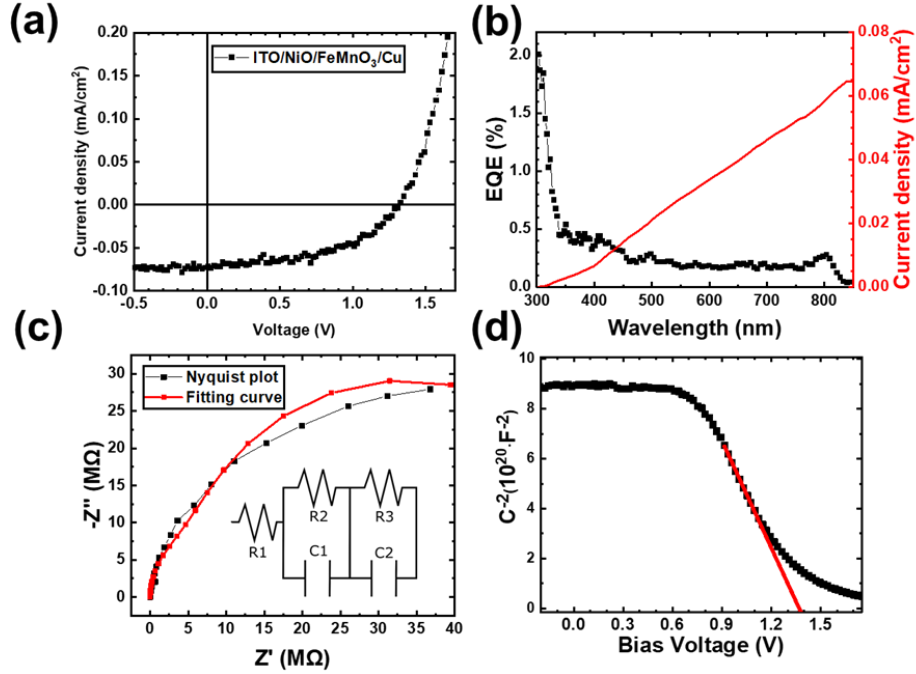

**Fig. 5** (a) J-V plot under illumination conditions for the representative p-n device under this study ITO/NiO/FeMnO$_3$/Cu. (b) EQE spectrum of NiO−FeMnO$_3$ (p−n) heterojunction sandwiched between ITO and Cu electrodes. The right axis represents the integrated photocurrent density of the corresponding device. (c) Nyquist, and (d) Mott–Schottky plots for the ITO/NiO/ FeMnO$_3$/Cu device.

To obtain further insights on the depletion regions in p−n devices under study and to further understand the charge dynamic processes, we performed EIS measurements under solar light irradiation and zero bias. Fig. 5c shows characteristic Nyquist plots of the NiO/FeMnO$_3$ heterojunction structure as well as an equivalent circuit model used to fit the experimental data; the components $R_1$, $R_3$ and $R_3$ can be ascribed to the contacting, charge transfer and recombination resistance, respectively.[79] The obtained results are presented in Table 2. Comparing with different types of metal oxide solar cells we can infer that the ITO/NiO/FeMnO$_3$/Cu device exhibits a relative high recombination resistance which explains the high $V_{oc}$ value, but on the other hand shows a relative high charge transport resistance which results in limited current density.[80] In Fig. 5d, the Mott-Schottky measurements of the title device which was swept from low to high external applied bias are illustrated. According to Mott-Schottky analysis the crossing of extrapolated linear section of the spectra with x-axis can be ascribed to the built-in potential of the device. Assuming that the proposed device is fully depleted during the measurement we exact a built-in potential of 1.38 V.[79]

**Table 2** Parameters obtained by fitting the Nyquist plots of the ITO/NiO/FeMnO$_3$/Cu device

| Solar Cell | $R_1$ (KΩ) | $R_2$ (MΩ) | $C_1$ (nF) | $R_3$ (MΩ) | $C_2$ (nF) |
|---|---|---|---|---|---|
| ITO/NiO/FMO/Cu | 1.83 | 5.52 | 0.053 | 57.27 | 0.059 |

Overall, to achieve higher efficiencies from FeMnO$_3$-based solar cells, further research is needed. Incorporation of buffer layers within the ITO/NiO/FeMnO$_3$/Cu device structure can be used to improved charge carrier selectivity and thus FF efficiency, while improvement of FeMnO$_3$ charge transport properties and thickness optimization are required to decrease the charge carrier recombination. Moreover, as indicated from Fig. 4b, in addition to the limited light harvesting capability of NiO the VB edge positions of p-type NiO and n-type FeMnO$_3$ materials are not well matched. Thus, the use of another p-type electronic material with suitable band edge alignment and lower optical bandgap could facilitate the holes transfer from FeMnO$_3$ to NiO and eventually improves the light harvesting PV capabilities. The above methods could provide increased photocurrent and thus resulting to a higher PCE FeMnO$_3$-based solar cells.

## 4. Conclusions

This study successfully proves that the synthesis of FeMnO$_3$ NPs can be achieved by a solution combustion technique, using tartaric acid as a fuel. Furthermore, the ultimate control of the nanoparticles' size can be readily attained due to the multiple binding ability of the tartaric acid that resulted in the formation of single-phase FeMnO$_3$ with an average particle size of 13 ± 2 nm. The high phase purity and crystallinity of FeMnO$_3$ were verified by X-ray diffraction and electron microscopy measurements. Additionally, we used a method to construct a network of tightly connected FeMnO$_3$ nanoparticles by spin coating of the colloidal solution. In this study, $β$-alanine was used as surface capping agent to produce a stable colloidal dispersion of FeMnO$_3$ NPs ($β$-alanine-capped FeMnO$_3$ NPs) in DMF. The short chain length of $β$-alanine allows direct interactions between the nanoparticles through ligand removal by thermal annealing (at 450 ºC in air), thus yielding a thick absorbing film (~500 nm) consisting of continuous layers of interconnected nanoparticles and exhibiting a relative low roughness of ~27 nm. The proposed strategy is crucial to obtain functional all-oxide photovoltaic devices that are fabricated by a simple solution process technique. Furthermore, for the first time, the inorganic perovskite FeMnO$_3$ was tested as a light absorber for photovoltaic applications. The band gap (~1.54 eV) of the synthesized FeMnO$_3$ nanostructure was found to be very close to the hybrid lead perovskite CH$_3$NH$_3$PbI$_3$ material (~1.55 eV). To this end, all-inorganic NiO/FeMnO$_3$ heterojunction photovoltaics were fabricated by solution combustion synthesis, using spin coating techniques. The corresponding all-inorganic solar cells reveal a high open circuit voltage ($V_{oc}$) of 1.31 V with a fill factor (FF) of 54.3% but exhibit a PCE of 0.05% under 100 mW cm$^{-2}$ illumination due to the limited short circuit current 0.07 mA cm$^{-2}$. Electrical characterization by impedance spectroscopy showed that the ITO/NiO/FeMnO$_3$/Cu device exhibits a high recombination resistance justifying the high $V_{oc}$. The high charge transport resistance indicates charge transport limitations within the relative thick (~500 nm) n-type FeMnO$_3$ active layer. Optimizing the active layer thickness, improving the charge carrier transport properties of n-type FeMnO$_3$ and replacing NiO with another p-type material with suitable optical band gap would improve $J_{sc}$ values. Moreover, the incorporation of suitable hole and electron transporting layers within the solar cell device architecture will improve electrodes charge carrier selectivity and FF values. The obtained results encourage a more intense research on solution processed and environmentally friendly inorganic solar cells with suitable opto-electronic properties and high photon to electron conversion efficiency.

**Acknowledgements**

This research was funded by the European Research Council (ERC) under the European Union's Horizon 2020 research and innovation program (grant agreement No 647311) and further supported from the academic yearly research activity internal Cyprus University of Technology budget.

**References**

1   O. Ellabban, H. Abu-Rub and F. Blaabjerg, *Renew. Sustain. Energy Rev.*, 2014, **39**, 748–764.

2   A. K. Hussein, *Renew. Sustain. Energy Rev.*, 2015, **42**, 460–476.

3   W. Wang, M. T. Winkler, O. Gunawan, T. Gokmen, T. K. Todorov, Y. Zhu and D. B. Mitzi, *Adv. Energy Mater.*, 2014, **4**, 1–5.

4   D. B. Mitzi, O. Gunawan, T. K. Todorov, K. Wang and S. Guha, *Sol. Energy Mater. Sol. Cells*, 2011, **95**, 1421–1436.

5   E. D. Kosten, J. H. Atwater, J. Parsons, A. Polman and H. A. Atwater, *Light Sci. Appl.*, 2013, **2**, e45.

6   A. Rohatgi, S. Narasimha, S. Kamra, P. Doshi, C. P. Khattak, K. Emery and H. Field, in *Conference Record of the Twenty Fifth IEEE Photovoltaic Specialists Conference - 1996*, IEEE, 1996, 741–744.

7   E. H. Jung, N. J. Jeon, E. Y. Park, C. S. Moon, T. J. Shin, T.-Y. Yang, J. H. Noh and J. Seo, *Nature*, 2019, **567**, 511–515.

8   F. Li and M. Liu, *J. Mater. Chem. A*, 2017, **5**, 15447–15459.

9   Q. Fu, X. Tang, B. Huang, T. Hu, L. Tan, L. Chen and Y. Chen, *Adv. Sci.*, 2018, **5**, 1700387.

10  A. Pérez-Tomás, A. Mingorance, D. Tanenbaum and M. Lira-Cantú, *Metal Oxides in Photovoltaics: All-Oxide, Ferroic, and Perovskite Solar Cells*, Elsevier Inc., 2018.

11  H. Kim, C. M. Gilmore, A. Piqué, J. S. Horwitz, H. Mattoussi, H. Murata, Z. H. Kafafi and D. B. Chrisey, *J. Appl. Phys.*, 1999, **86**, 6451–6461.

12  C.-H. Han, S.-D. Han, J. Gwak and S. P. Khatkar, *Mater. Lett.*, 2007, **61**, 1701–1703.

13  J.-Y. Seo, R. Uchida, H.-S. Kim, Y. Saygili, J. Luo, C. Moore, J. Kerrod, A. Wagstaff, M. Eklund, R. McIntyre, N. Pellet, S. M. Zakeeruddin, A. Hagfeldt and M. Grätzel, *Adv. Funct. Mater.*, 2018, **28**, 1705763.

14  S. S. Shin, S. J. Lee and S. I. Seok, *APL Mater.*, 2019, **7**, 022401.

15  L. Xiong, Y. Guo, J. Wen, H. Liu, G. Yang, P. Qin and G. Fang, *Adv. Funct. Mater.*, 2018, **28**, 1802757.

16  I. T. Papadas, F. Galatopoulos, G. S. Armatas, N. Tessler and S. A. Choulis, *Nanomaterials*, 2019, **9**, 1616.

17  F. Galatopoulos, A. Savva, I. T. Papadas and S. A. Choulis, *APL Mater.*, 2017, **5**, 076102.

18  I. T. Papadas, A. Savva, A. Ioakeimidis, P. Eleftheriou, G. S. Armatas and S. A. Choulis, *Mater. Today Energy*, 2018, **8**, 57–64.

19  I. T. Papadas, A. Ioakeimidis, G. S. Armatas and S. A. Choulis, *Adv. Sci.*, 2018, **5**, 1–9.

20  A. Ioakeimidis, I. T. Papadas, D. Tsikritzis, G. S. Armatas, S. Kennou and S. A. Choulis, *APL Mater.*, 2019, **7**, 021101.

21  A. Pérez-Tomas, H. Xie, Z. Wang, H. S. Kim, I. Shirley, S. H. Turren-Cruz, A. Morales-Melgares, B. Saliba, D. Tanenbaum, M. Saliba, S. M. Zakeeruddin, M. Gratzel, A. Hagfeldt and M. Lira-Cantu, *Sustain. Energy Fuels*, 2019, **3**, 382–389.

22  I. Grinberg, D. V. West, M. Torres, G. Gou, D. M. Stein, L. Wu, G. Chen, E. M. Gallo, A. R. Akbashev, P. K. Davies, J. E. Spanier and A. M. Rappe, *Nature*, 2013, **503**, 509–512.

23  I. Papadas, J. A. Christodoulides, G. Kioseoglou and G. S. Armatas, *J. Mater. Chem. A*, 2015, **3**, 1587–1593.

24  R. Nechache, C. Harnagea, S. Li, L. Cardenas, W. Huang, J. Chakrabartty and F. Rosei, *Nat. Photonics*, 2014, **9**, 61–67.

25  Y. Sun, F. Guo, J. Chen and S. Zhao, *Appl. Phys. Lett.*, 2017, **111**, 253901.

26  S. Calnan, *Coatings*, 2014, **4**, 162–202.

27  G. Zhang, H. Wu, G. Li, Q. Huang, C. Yang, F. Huang, F. Liao and J. Lin, *Sci. Rep.*, 2013, **3**, 1265.

28  K. C. Patil, S. T. Aruna and T. Mimani, *Curr. Opin. Solid State Mater. Sci.*, 2002, **6**, 507–512.

29  T. Mimani and K. C. Patil, *Mater. Phys. Mech.*, 2001, **4**, 134–137.

30  N. P. Bansal, Springer Science & Business Media, 2006, **200**, pp 554.

31  K. Suresh and K. C. Patil, 1991, **560012**, 148–150.

32  F. Deganello, G. Marcì and G. Deganello, *J. Eur. Ceram. Soc.*, 2009, **29**, 439–450.

33  L. D. Jadhav, S. P. Patil, A. U. Chavan, A. P. Jamale and V. R. Puri, *Micro & Nano Lett.*, 2011, **6**, 812–815.

34  Q. Jiang, J. Lu, J. Cheng, X. Li, R. Sun, L. Feng, W. Dai, W. Yan and Z. Ye, *Appl. Phys. Lett.*, 2014, **105**, 132105.

35  M.-G. Kim, M. G. Kanatzidis, A. Facchetti and T. J. Marks, *Nat. Mater.*, 2011, **10**, 382–388.

36 H. H. Hsieh, X. Yu, Y. Xia, C. W. Sheets, C. C. Hsiao, T. J. Marks and A. A. Facchetti, *Dig. Tech. Pap. - SID Int. Symp.*, 2014, **45**, 427–430.

37 X. Yu, J. Smith, N. Zhou, L. Zeng, P. Guo, Y. Xia, A. Alvarez, S. Aghion, H. Lin, J. Yu, R. P. H. Chang, M. J. Bedzyk, R. Ferragut, T. J. Marks and A. Facchetti, *Proc. Natl. Acad. Sci. U. S. A.*, 2015, **112**, 3217–22.

38 M. H. Habibi and V. Mosavi, *J. Mater. Sci. Mater. Electron.*, 2017, **28**, 8473–8479.

39 K. Cao, H. Liu, X. Xu, Y. Wang and L. Jiao, *Chem. Commun.*, 2016, **52**, 11414–11417.

40 A. Cetin, A. M. Önal and E. N. Esenturk, 2019, **257**, 1–9.

41 C. Doroftei, P. Dorin, E. Rezlescu and N. Rezlescu, *Compos. PART B*, 2014, **67**, 179–182.

42 Z. Z. Vasiljevic, M. P. Dojcinovic, J. B. Krstic and V. Ribic, 2020, **10**, 13879–13888.

43 M. V. Nikolic, J. B. Krstic, N. J. Labus, M. D. Lukovic, M. P. Dojcinovic, M. Radovanovic and N. B. Tadic, *Mater. Sci. Eng. B Solid-State Mater. Adv. Technol.*, 2020, **257**, 114547.

44 B. Saravanakumar, S.P. Ramachandran, G. Ravi, V. Ganesh, Ramesh K. Guduru, R. Yuvakkumar, *Materials Research Express*, 2018, **5**, 015504.

45 D. Soni and R. Pal, *Electroanalysis*, 2016, **28**, 1951–1956.

46 T. Fix, *Advanced Micro- and Nanomaterials for Photovoltaics*, 2019, **2**, 19-34.

47 S. Gowreesan and A. Ruban Kumar, *Appl. Phys. A Mater. Sci. Process.*, 2017, **123**, 1–8.

48 P. B. Mungse, G. Saravanan, M. Nishibori and J. Subrt and N. K. Labhsetwar, *Pure and App. Chem.*, 2017, **89**, 511-521.

49 H. Bin, Z. Yao, S. Zhu, C. Zhu, H. Pan, Z. Chen, C. Wolverton and D. Zhang, *J. Alloys Compd.*, 2017, **695**, 1223–1230.

50 R. Sundari, T. I. Hua and M. Rusli Yosfiah, *Adv. Mater. Res.*, 2013, **634–638**, 2150–2154.

51 L. S. Lobo and A. Rubankumar, *Ionics (Kiel).*, 2019, **25**, 1341–1350.

52 E. Skliri, J. Miao, J. Xie, G. Liu, T. Salim, B. Liu, Q. Zhang and G. S. Armatas, *Appl. Catal. B Environ.*, 2018, **227**, 330–339.

53 J. Bennet, R. Tholkappiyan, K. Vishista, N. V. Jaya and F. Hamed, *Appl. Surf. Sci.*, 2016, **383**, 113–125.

54 A. Alves, C. P. Bergmann, F. A. Berutti, Novel Synthesis and Characterization of Nanostructured Materials, *Springer*, 2013, ISBN 978-3-642-41275-2.

55 R. Buonsanti, T. E. Pick, N. Krins, T. J. Richardson, B. A. Helms and D. J. Milliron, *Nano Lett.*, 2012, **12**, 3872–3877.

56 A. Savva, I. Papadas, D. Tsikritzis, G. S. Armatas, S. Kennou and S. Choulis, *J. Mater. Chem. A*, 2017, **5**, 20381–20389.

57    S. Kerli and Alver, *J. Nanotechnol.*, 2016, **2016**, 1-5.

58    T. Mahmood, M. T. Saddique, A. Naeem, P. Westerhoff, S. Mustafa and A. Alum, *Ind. Eng. Chem. Res.*, 2011, **50**, 10017–10023.

59    H. Zhang, J. Cheng, F. Lin, H. He, J. Mao, K. S. Wong, A. K.-Y. Jen and W. C. H. Choy, *ACS Nano*, 2015, **9**, 639-646.

60    A. Chrissanthopoulos, S. Baskoutas, N. Bouropoulos, V. Dracopoulos, P. Poulopoulos and S. N. Yannopoulos, *Photonics Nanostructures - Fundam. Appl.*, 2011, **9**, 132–139.

61    I. T. Papadas, K. S. Subrahmanyam, M. G. Kanatzidis and G. S. Armatas, *Nanoscale*, 2015, **7**, 5737–5743.

62    P. Kubelka, *J. Opt. Soc. Am.*, 1948, **38**, 448.

63    E. A. Cochran, K. N. Woods, D. W. Johnson, C. J. Page and S. W. Boettcher, *J. Mater. Chem. A*, 2019, **7**, 24124–24149.

64    L. A. Chick, L. R. Pederson, G. D. Maupin, J. L. Bates, L. E. Thomas and G. J. Exarhos, *Mater. Lett.*, 1990, **10**, 6–12.

65    K. Rajeshwar and N. R. de Tacconi, *Chem Soc Rev*, 2009, **38**, 1984–1998.

66    S. Verma, H. M. Joshi, T. Jagadale, A. Chawla, R. Chandra and S. Ogale, *J. Phys. Chem. C*, 2008, **112**, 15106–15112.

67    S. M. Selbach, M.-A. Einarsrud, T. Tybell and T. Grande, *J. Am. Ceram. Soc.*, 2007, **90**, 3430–3434.

68    Y. Wang, Y. Hu, L. Fei, Y. Zhang, J. Yuan and H. Gu, *J. Nanomater.*, 2011, **2011**, 1-6.

69    K. C. Patil, S. T. Aruna and T. Mimani, *Curr. Opin. Solid State Mater. Sci.*, 2002, **6**, 507–512.

70    M. Li, W. Xu, W. Wang, Y. Liu, B. Cui and X. Guo, *J. Power Sources*, 2014, **248**, 465–473.

71    M. A. Haque, A. D. Sheikh, X. Guan and T. Wu, *Adv. Energy Mater.*, 2017, **7**, 1602803.

72    K. M. Batoo, S. Kumar, C. G. Lee and Alimuddin, *Curr. Appl. Phys.*, 2009, **9**, 1397–1406.

73    E. Veena Gopalan, K. A. Malini, S. Saravanan, D. Sakthi Kumar, Y. Yoshida and M. R. Anantharaman, *J. Phys. D. Appl. Phys.*, 2008, **41**, 185005.

74    M. J. Akhtar and M. Younas, *Solid State Sci.*, 2012, **14**, 1536–1542.

75    N. Rezlescu and E. Rezlescu, *Phys. Status Solidi*, 1974, **23**, 575–582.

76    S. Chatterjee, A. Bera and A. J. Pal, *ACS Appl. Mater. Interfaces*, 2014, **6**, 20479–20486.

77    Y. H. Paik, H. S. Kojori, J.-H. Yun and S. J. Kim, *Mater. Lett.*, 2016, **185**, 247–251.

78    L. J. Tang, X. Chen, T. Y. Wen, S. Yang, J. J. Zhao, H. W. Qiao, Y. Hou and H. G. Yang, *Chem. - A Eur. J.*, 2018, **24**, 2845–2849.

79  E. V. Hauff, *J. Phys. Chem. C*, 2019, **123**, 11329–11346.

80  M. Sariful Sheikh, D. Ghosh, A. Dutta, S. Bhattacharyya and T. P. Sinha, *Mater. Sci. Eng. B Solid-State Mater. Adv. Technol.*, 2017, **226**, 10–17.

# Supporting information

## All-Inorganic p−n Heterojunction Solar Cells by Solution Combustion Synthesis using n-type FeMnO$_3$ Perovskite Photoactive Layer

**Ioannis T. Papadas,[a,c]\*  Apostolos Ioakeimidis,[a] Ioannis Vamvasakis,[b] Polyvios Eleftheriou [a] Gerasimos S. Armatas[b] and Stelios A. Choulis[a]\***

[a] Molecular Electronics and Photonics Research Unit, Department of Mechanical Engineering and Materials Science and Engineering, Cyprus University of Technology, Limassol, Cyprus.
[b] Department of Materials Science and Technology, University of Crete, Heraklion 70013, Greece.
[c] Department of Public and Community Health, School of Public Health, University of West Attica, Athens, Greece

**\* Correspondence:**
Corresponding Authors: Assistant Prof. Ioannis T. Papadas, Prof. Stelios A. Choulis
emails: ioannis.papadas@cut.ac.cy, stelios.choulis@cut.ac.cy

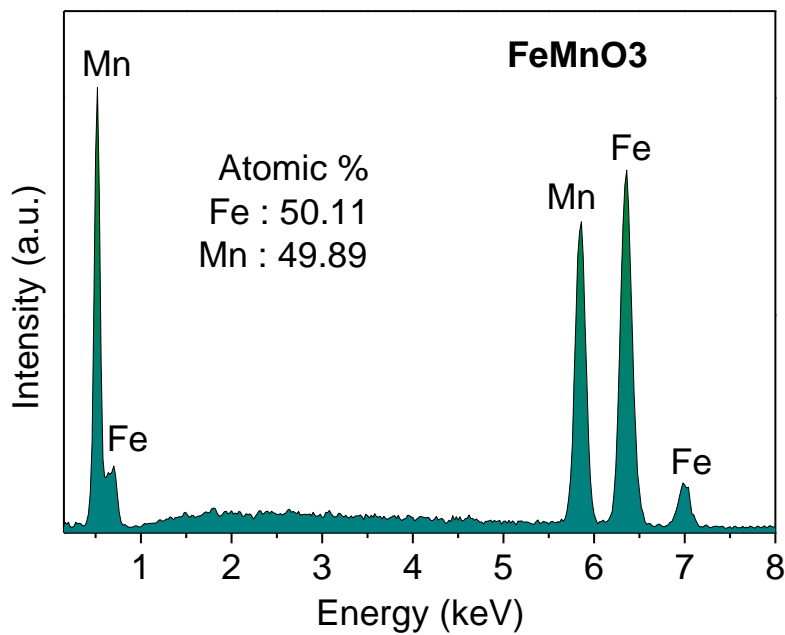

**Fig. S1.** Typical EDS spectrum for FeMnO$_3$ nanoparticles. The EDS analysis indicates an average atomic proportion of Mn:Fe ~1:1.

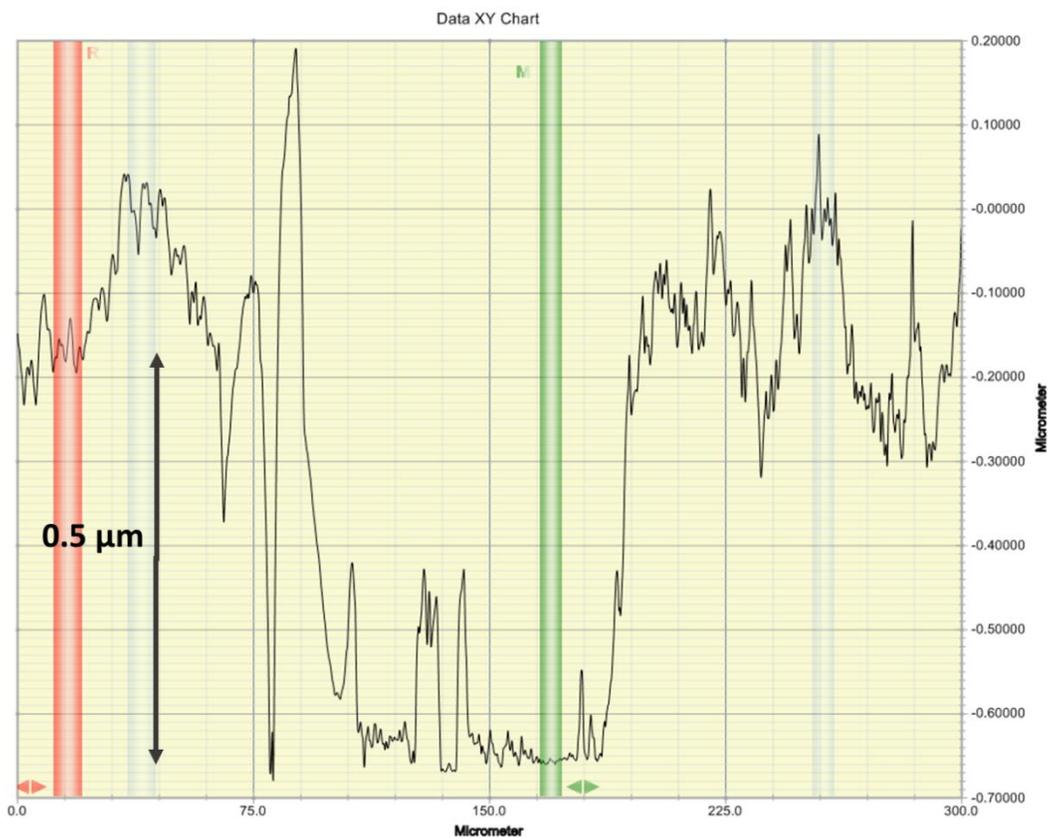

**Fig. S2.** Surface profile measurements of FeMnO$_3$ nanoparticles thick layer (~500 nm) under study.

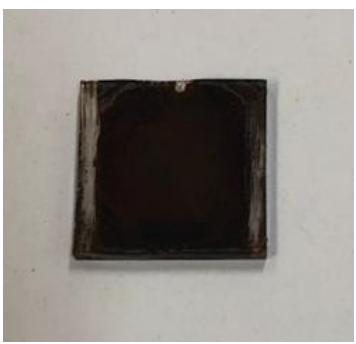

**Fig. S3.** Photograph of FeMnO$_3$ nanoparticles film obtained by spin coating technique.